\documentstyle[preprint,eqsecnum,aps]{revtex}
\begin{document}
\draft
\title{\hfill{\small{DSF Preprint 95/50, gr-qc/9512033}}\\
Coleman-Weinberg SO(10) GUT Theories as Inflationary
Models}
\author{Giampiero Esposito, \thanks{Electronic address:
esposito@napoli.infn.it} Gennaro Miele, \thanks{Electronic
address: miele@napoli.infn.it}
and Pietro Santorelli \thanks{Electronic address:
santorelli@napoli.infn.it}}
\address{Istituto Nazionale di Fisica Nucleare, Sezione di
Napoli,\\
Mostra d'Oltremare Padiglione 20, 80125 Napoli, Italy \\
Dipartimento di Scienze Fisiche, \\
Mostra d'Oltremare Padiglione 19, 80125 Napoli, Italy \\}
\maketitle
\vspace{-1.truecm}
\begin{abstract}
The flat-space limit of the one-loop effective potential for
SO(10) GUT theories in spatially flat Friedmann-Robertson-Walker
cosmologies is applied to study the dynamics of the early
universe. The numerical integration of the corresponding field equations
shows that, for such grand unified theories, a sufficiently long
inflationary stage is achieved for suitable choices of the
initial conditions. However, a severe fine tuning of these initial
conditions is necessary to obtain a large $e$-fold number.
In the direction with residual symmetry $SU(4)_{PS} \otimes
SU(2)_{L} \otimes SU(2)_{R}$, one eventually finds parametric
resonance for suitable choices of the free parameters
of the classical potential.
This phenomenon leads in turn to the end of inflation.
\end{abstract}

\section {Introduction}
Semiclassical effects in quantum field theory are at the very
heart of many exciting developments in modern theoretical
physics, e.g., radiative corrections to the Casimir force [1],
trace anomalies [2], one-loop effective
action [3,4], symmetry breaking
in the early universe [5--7]. In the latter class of
phenomena, the one-loop effective potential for grand unified
theories (GUT) in curved backgrounds is used in the field equations
to determine the symmetry-breaking pattern which is relevant
for the familiar low-energy phenomenology of particle
physics [6,7]. In particular,
in a previous paper by the first two
authors, the one-loop effective
potential for SO(10) GUT theories in de Sitter space
was derived for the first time [8]. The main motivation
of our work was the analysis of spontaneous symmetry breaking
in the early universe when GUT theories in good agreement
with low-energy phenomenology are studied. Unlike $SU(5)$
theories, $SO(10)$ models make it possible to allocate a
whole fermionic family in the single irreducible spinorial
representation, and predict longer nucleon lifetimes in
agreement with experimental data. This happens because
lepto-quarks responsible for nucleon decays have higher masses
(with respect to $SU(5)$), by virtue of an intermediate
symmetry group. The latter lies in between the highest scale
$M_{X} \cong 10^{16}$ GeV, and the electroweak scale. Further
details can be found in Refs. [8,9].

On considering the particular mass matrix relevant for
the $SU(3)_{C} \otimes SU(2)_{L} \otimes U(1)_{Y}$ symmetry-breaking
direction, our main result was the numerical proof that,
even when curvature effects are no longer negligible, the early
universe can only reach the $SU(4)_{PS} \otimes SU(2)_{L}
\otimes SU(2)_{R}$ absolute minimum [8].

A naturally occurring
question is whether the analytic formulas obtained in Ref. [8] can
be used to gain a better understanding of other phenomena
occurring in the early universe. Indeed, it is well-known
that, according to the cosmological standard model based
on Einstein's general relativity,
the early universe is spatially homogeneous and isotropic [10,11],
and gravity couples to the energy-momentum tensor of matter.
Under the previous assumptions of symmetry and homogeneity
(cf. Ref. [12]), the
space-time metric can be locally cast in the form [10]
\begin{equation}
g=-dt \otimes dt + a^{2}(t)
\biggr [d\chi \otimes d\chi +f^{2}(\chi)
\Bigr(d\theta_{1} \otimes d\theta_{1}+\sin^{2}{\theta_{1}}~
d\theta_{2} \otimes d\theta_{2}\Bigr) \biggr] \; ,
\end{equation}
where $a(t)$ is the cosmic scale factor,
$\theta_{1} \in ]0,\pi[$, $\theta_{2} \in ]0,2\pi[$
and, denoting by $k$
the constant curvature of the three-dimensional spatial sections,
$f(\chi)=\sin \chi, \chi$ or $\sinh \chi$ if $k=1,0,-1$
respectively. Here we focus on spatially
flat Friedmann-Robertson-Walker (FRW) cosmologies, for which $k=0$.

The aim of the present work has been therefore to study the
relevance of SO(10) GUT models for gravitational physics
within the more general framework of FRW cosmologies. In this
case, the analysis performed in
Ref. [13] enables one to use, at least for
typical GUT-inflationary models characterized by a phase-transition
temperature $\leq 10^{16}~GeV$, the flat-space limit of
the Coleman-Weinberg one-loop effective potential
[5]. What happens is that the one-loop field equations in
FRW cosmologies are an involved set of integro-differential
equations [13]. In scalar electrodynamics,
the nonlocal correction to the Coleman-Weinberg
effective potential results from the coupling of a scalar field
$\phi_{c}$ to the gravitational background and may
lead to dissipative effects in the early universe. More
precisely, nonlocal terms in the field equations may be
approximated very well by a function of the form [13]
$$
-A(\tilde \tau) \biggr(Ha \phi_{c}
+{d\phi_{c}\over d{\tilde \tau}}\biggr)
{d\phi_{c}\over d{\tilde \tau}}
$$
where $A$ is a positive-definite
function of the conformal time $\tilde \tau$.
The first term in round brackets leads to a further quantum
correction to the energy density of the scalar field (for
slowly varying $A$), while the second term is purely
dissipative [13]. However, nonlocal terms {\it do not}
modify the pattern of minima found in static de Sitter space.
Thus, for numerical purposes, even in
the nonAbelian case, curvature effects on the
one-loop potential are very small, and can be neglected at
the GUT energy scale. This is the basic approximation on
which the present paper relies.

In Sec. II the scalar dynamics
driving the inflationary phase is described. Section III,
relying on Ref. [8], studies SO(10) GUT models with Higgs field
in the 210-dimensional irreducible representation.
In the $SO(10)$ internal space, spherical coordinates for the Higgs
field are used to obtain a convenient parametrization of the
classical part of the potential. The
flat-space limit of the one-loop effective
potential is then studied
in Sec. IV. Semiclassical field equations are studied in
Sec. V, and their numerical solution is found in Sec. VI.
The reheating process is then briefly described in Sec. VII,
while the concluding remarks are presented in Sec. VIII.

\section{Inflation from scalar-field dynamics}

In the inflationary models driven by scalar fields,
the universe contains
different species of matter, i.e., radiation and scalar particles
(inflatons). The latter, characterized by a higher energy level
with respect to the others, completely determine the dynamics of
the universe, making all remaining contributions negligible.
A theoretical scheme in which fundamental scalar particles naturally occur
(Higgs field) and lead to symmetry breaking
are the GUT theories. In this case the
energies involved are compatible with the
levels required for an {\it efficient} inflation. Thus, we consider
an inflationary model in which the inflaton coincides with
the Higgs field of a GUT theory, corresponding
to the highest energy scale of spontaneous symmetry breaking.
In particular, we choose the most relevant ones which are the SO(10) GUT
theories with Higgs field belonging to
the $210$-dimensional irreducible representation
($\underline{210}$) [8].

Within this framework (we denote hereafter by $\phi$ the fields
corresponding to scalar particles), and assuming homogeneous field
configurations $\phi=\phi(t)$, the
{\it semiclassical} Lagrangian reads
\begin{equation}
{\cal L} = 6 \dot{a}^2a F(\phi)+6 \dot{a} a^2 F'(\phi) \dot{\phi}
+ a^3\left[ { 1 \over 2} \dot{\phi}^2 -V^{(1)}(\phi)\right] \;,
\end{equation}
where the {\it dot} stands for $d/dt$,
the {\it prime} for $\delta/ \delta\phi$
and $F(\phi)$ denotes the arbitrary coupling
of $\phi$ to the gravitational background with scalar curvature
$R=6\left[{{\ddot a}\over a}
+{{\dot a}^{2}\over a^{2}}\right]$.
In Eq. (2.1) the classical potential has been replaced by
the one-loop effective potential $V^{(1)}(\phi)$,
i.e., the gauge degrees of freedom
have been integrated over.
Note also that the first two terms in (2.1)
yield the coupling $-a^{3}F(\phi) R$,
after integration by parts in the action.

Starting from (2.1) one gets the two
semiclassical equations of motion
\begin{equation}
{\ddot{a} \over a} + { 1\over 2} \left( {\dot{a} \over a}\right)^2+
\left[ { \dot{a} \over a} { F' \over F} \dot{\phi} +
{1\over 2} {F' \over F} \ddot{\phi}
+ { 1 \over 2} {F'' \over F} \dot{\phi}^2\right]
- { 1 \over 8} {\dot{\phi}^2  \over F} + { 1 \over 4}
{V^{(1)} \over F}=0 \; ,
\end{equation}
and
\begin{equation}
\ddot{\phi} + 3 {\dot{a} \over a} \dot{\phi} + 6
\left[{{\ddot a}\over a}
+ {{\dot a}^{2}\over a^{2}}\right] F'(\phi)+{\delta V^{(1)} \over \delta \phi}
=0 \;.
\end{equation}
In general, for a nonminimally coupled
real scalar field, $F(\phi) = (1/16 \pi G)
+ (\xi/2) \phi^2$. In our case,
in the light of the results obtained in Ref. [13],
we neglect all contributions of the
curvature to the scalar potential,
and hence we can assume minimally coupled scalar
fields (for which $\xi$ vanishes).

\section{SO(10) gut models with higgs field in the \protect\\
$\underline{210}$ irreducible representation}

In a gauge unifying theory like $SO(10)$, the Higgs field (fundamental
scalar particles) belongs to one or more
irreducible representations (hereafter referred to as
IRR's) of the gauge group, and its
dynamics is ruled by a Higgs potential.
These particles provide the correct residual symmetry for the
model in the low-energy limit, through a spontaneous
symmetry-breaking mechanism.
To study the inflation corresponding to the highest
energy scale ($M_{X}$) of spontaneous
symmetry breaking (SSB), we can consider
the only contribution of the IRR
responsible for the SSB at that scale.

In the present case, we
consider the most general renormalizable Higgs potential constructed
by using a massless and minimally coupled
IRR $\underline{210}$ only, which is obtained by
the completely anti-symmetrized product
of four different $\underline{10}$'s as
\begin{equation}
\Phi_{abcd}=N \; \mu_{[a}\otimes
\nu_{b}\otimes\rho_{c}\otimes
\sigma_{d]} \; ,
\end{equation}
where $N$ is a normalization constant.
The $\underline{210}$ IRR has four independent quartic
invariants, i.e., ${\| \phi \|}^{4}$ and three non-trivial
invariants, hence the Higgs potential we are going to construct
is a function of these [8]
\begin{equation}
V(\phi) =  V_0 + g_{1}~\|(\phi \phi)_{\underline {45}}\|^{2}
+ g_{2}~\|(\phi \phi)_{\underline {210}}\|^{2}+
g_{3}~\|(\phi \phi)_{\underline {1050}}\|^{2}
+\lambda~ \|\phi\|^{4} \; ,
\end{equation}
where $g_{1}$, $g_{2}$, $g_{3}$ and $\lambda$ are
arbitrary coefficients, and $V_0$ is an arbitrary constant.
We will see in due course what is the meaning of this
constant and how it can be fixed.
Unfortunately, in view of the technical difficulties
to express (3.2) in terms of the $210$
degrees of freedom, we restrict our analysis to the only
directions invariant under the subgroup
$SU(3)_{C} \otimes SU(2)_{L}
\otimes U(1)_{Y}$. This restriction, however,
remains relevant for the aims of this paper, since
it leads to the correct electroweak phenomenology
at low energies for the model.
The most general singlet $\phi_{0}$ with respect to the group
$SU(3)_{C} \otimes SU(2)_{L} \otimes U(1)_{Y}$
contained in the $\underline{210}$ representation is [8]
\begin{eqnarray}
{\phi_{0}\over \| \phi_{0} \|}&=&{1\over \sqrt{6}}
\sin{\theta} \sin{\varphi}
\Bigr(\phi_{1278}+\phi_{3478}+\phi_{5678}+
\phi_{1290}+\phi_{3490}+\phi_{5690}\Bigr) \nonumber \\
&+&{1\over \sqrt{3}}
\sin{\theta} \cos{\varphi}
\Bigr(\phi_{1234}+\phi_{3456}+
\phi_{5612}\Bigr)+ \cos{\theta} \Bigr(\phi_{7890}\Bigr) \; ,
\end{eqnarray}
where $\theta \in [0, \pi]$ and $\varphi \in [0, 2 \pi[$. In particular,
by varying $\theta$ and $\varphi$ in
their ranges, one gets the following residual-symmetry groups [8]:
\begin{mathletters}
\begin{equation}
\theta=0,\pi~~~ \mbox {and/or}~~~\varphi=0,\pi
\rightarrow  SU(3)_{C} \otimes SU(2)_{L}
\otimes SU(2)_{R} \otimes U(1)_{B-L} \; ,
\end{equation}
\begin{equation}
\theta={\pi \over 2}~~~ \mbox {and} ~~~\varphi={0,\pi}
\rightarrow SU(3)_{C} \otimes SU(2)_{L}
\otimes SU(2)_{R} \otimes U(1)_{B-L} \times D \; ,
\end{equation}
\begin{equation}
\theta=0,\pi \rightarrow SU(4)_{PS} \otimes SU(2)_{L} \otimes
SU(2)_{R} \; ,
\end{equation}
\begin{equation}
\theta=\arctan(3)~~~ \mbox {and}~~~\varphi
=\arctan(\sqrt{2})~~~ \mbox {or}
\end{equation}
\begin{equation}
\theta=-\arctan(3) +\pi~~~
\mbox {and}~~~\varphi=\arctan(\sqrt{2})+\pi
\rightarrow SU(5) \otimes U(1) \; ,
\end{equation}
\begin{equation}
\mbox {otherwise} \rightarrow SU(3)_{C} \otimes SU(2)_{L}
\otimes U(1)_{T_{3R}} \otimes U(1)_{B-L} \; ,
\end{equation}
\end{mathletters}
where $T_{3R}$ is the $z$-component of the $SU(2)_R$ group.

Inserting (3.3) into (3.2) one finds the tree-level potential
\begin{equation}
V(\phi_{0}) =  V_0 + \biggr({\alpha \over 8}f_{\alpha}
+{\gamma \over 4}f_{\gamma}+{\delta \over 9}
f_{\delta} + (\lambda - \delta) \biggr)
\| \phi_{0} \|^{4} \; ,
\end{equation}
where [8]
\begin{equation}
\alpha \equiv {4\over 945}\Bigr(-108g_{1}+28g_{2}+140g_{3}\Bigr) \; ,
\end{equation}
\begin{equation}
\gamma \equiv {8\over 35}g_{1} \; ,
\end{equation}
\begin{equation}
\delta \equiv -{1\over 10}g_{3} \; ,
\end{equation}
\begin{equation}
f_{\alpha} \equiv \sin^4{\theta}
+\sin^2{\theta} \sin^2{\varphi} \left[
2 \sin{\theta} \cos{\varphi} + \sqrt{3} \cos{\theta} \right]^{2}
+ {3 \over 4} \sin^4{\theta} \sin^4{\varphi} \; ,
\end{equation}
\begin{equation}
f_{\gamma} \equiv \sin^2{\theta} \left[ \cos{\theta} \cos{\varphi}
+  {1\over \sqrt{3}}
\sin{\theta} \sin^2{\varphi} \right]^{2}+
\sin^4{\theta} \sin^2{\varphi} \cos^2{\varphi}
+f_{\alpha} \; ,
\end{equation}
\begin{equation}
f_{\delta} \equiv  \left[ 2 \sin^2{\theta} \cos^2{\varphi}
-{1\over 2} \sin^2{\theta} \sin^2{\varphi}
-3 \cos^2{\theta} \right]^{2} + 30 f_{\gamma}
-25 f_{\alpha} \; .
\end{equation}
Since in the following analysis
$\delta$ is always negative and $\alpha$ may take
negative values, the tree-level potential (3.5) is unbounded
from below, unless we impose the restriction [8]
\begin{equation}
\lambda \geq {\mid \alpha \mid \over 8}
\Bigr(f_{\alpha}\Bigr)_{{\rm max}}
+{\mid \delta \mid \over 9}
\Bigr(f_{\delta}\Bigr)_{{\rm max}} \; .
\end{equation}
Note also that contributions
proportional to a cubic term in the potential are set to zero,
since we are assuming $\phi \rightarrow -\phi$
invariance of our model.

\section{One-loop effective potential}

As shown by Coleman and Weinberg [5],
the effects of quantum fluctuations on the
scalar potential, due to both
gauge bosons and self-interactions,
may account for the spontaneous
symmetry breaking occurring in gauge theories
without assuming from the beginning the
presence of negative quadratic terms for the Higgs field.
These phenomena are best tackled in terms of the
one-loop effective potential [5--8,13,14]. This provides,
in our case, the appropriate tool for studying the phase
transition due to scalar dynamics.

The one-loop effective potential $V^{(1)}$ corresponding
to the classical expression (3.5) in the flat-space limit is obtained
in Ref. [8]. On defining
\begin{eqnarray}
h_{1}& \equiv &  \cos^2{\theta}+\sin^2{\theta} \left[{1\over 2}
\sin^2{\varphi}+{2\over 3} \cos^2{\varphi}\right] \nonumber \\
&+&\sqrt{2 \over 3} \sin{\theta} \sin{\varphi}
\left[ \cos{\theta} + { 2 \over \sqrt{3}} \sin{\theta} \cos{\varphi}
\right] \; ,
\end{eqnarray}
\begin{equation}
h_{2} \equiv  \cos^2{\theta}+\sin^2{\theta} \left[ {1 \over 2}
\sin^2{\varphi} +  {2\over 3} \cos^2{\varphi} \right] -
\sqrt{2\over 3} \sin{\theta}  \cos{\theta} \sin{\varphi} \; ,
\end{equation}
\begin{equation}
h_{3} \equiv {1\over 2} \sin^2{\theta} \sin^2{\varphi} \; ,
\end{equation}
\begin{equation}
h_{4} \equiv {2\over 3}  \sin^2{\theta} \; ,
\end{equation}
\begin{equation}
h_{5}^{2} \equiv {3\over 2}h_{1}^{2}+{3\over 2}h_{2}^{2}
+{h_{3}^{2}\over 4}+{3\over 4}h_{4}^{2} \; ,
\end{equation}
one gets
\begin{eqnarray}
V^{(1)}(\phi_{0})
& \sim & V_{0} +\biggr({\alpha \over 8}f_{\alpha}
+{\gamma \over 4}f_{\gamma}+{\delta \over 9}
f_{\delta} + (\lambda - \delta) \biggr) \| \phi_{0} \|^{4}
\nonumber \\
&-& {3 {\cal G}^4 \over 8\pi^{2}}
\| \phi_{0} \|^{4}
\biggr[{3\over 4}h_{1}^{2}\Bigr(3-\ln(h_{1}^{2})\Bigr)
+{3\over 4}h_{2}^{2}\Bigr(3-\ln(h_{2}^{2})\Bigr)
\nonumber \\
&+& {h_{3}^{2}\over 8}\Bigr(3-\ln(h_{3}^{2})\Bigr)
+{3\over 8}h_{4}^{2}\Bigr(3-\ln(h_{4}^{2})\Bigr)
-h_{5}^{2}\ln\left({ {\cal G}^2  \| \phi_{0} \|^{2} \over \mu^2}
\right)\biggr] \; ,
\end{eqnarray}
where ${\cal G}$ is the gauge coupling
constant and $\mu$ is the renormalization mass.

\section{Semiclassical field equations}

After defining the dimensionless time $\tau \equiv \mu t/{\cal G}$ the
Higgs field $\phi_{0}$ can be seen, in the $SO(10)$-space, as the
position vector in $R^3$, given in spherical coordinates as
$\tilde{\phi} \equiv ({\cal G}/\mu) \phi_{0} = y~\hat{e}_{y}$. Thus,
by taking its derivative with respect to $\tau$ one has
\begin{equation}
{d\over d\tau}
{\tilde{\phi} }= {dy\over d\tau} ~ \hat{e}_{y}
+ y ~ {d\theta \over d\tau} ~
\hat{e}_{\theta}  + y ~ \sin \theta~
{d\varphi \over d\tau}~ \hat{e}_{\varphi} \; .
\end{equation}
With our notation, it is convenient to introduce a
dimensionless expression for (4.6) obtained multiplying it by ${\cal
G}^4/\mu^4$, i.e.,
\begin{eqnarray}
{\widetilde V}^{(1)}(y,\theta,\varphi)
& \sim & \widetilde{V}_0 +\biggr({\alpha \over 8}f_{\alpha}
+{\gamma \over 4}f_{\gamma}+{\delta \over 9}
f_{\delta} + (\lambda - \delta) \biggr) y^{4}
\nonumber \\
&-&{3 {\cal G}^4 \over 8\pi^{2}}y^{4}
\biggr[{3\over 4}h_{1}^{2}\Bigr(3-\ln(h_{1}^{2})\Bigr)
+{3\over 4}h_{2}^{2}\Bigr(3-\ln(h_{2}^{2})\Bigr)
\nonumber \\
&+&{h_{3}^{2}\over 8}\Bigr(3-\ln(h_{3}^{2})\Bigr)
+{3\over 8}h_{4}^{2}\Bigr(3-\ln(h_{4}^{2})\Bigr)
-h_{5}^{2}\ln\left( y^{2} \right)\biggr] \; .
\end{eqnarray}
As far as the constant ${\widetilde V}_{0}$ is concerned, it has to be
fixed by requiring that a vanishing potential energy should correspond
to the absolute minimum for ${\widetilde V}^{(1)}$. By denoting with
$y_{m}$, $\theta_m$ and $\varphi_m$ the absolute-minimum coordinates,
we have $\widetilde{V}^{(1)}(y_{m},\theta_m,\varphi_m)$=0. Moreover,
$y_{m}$ makes it possible to determine also $\mu$, bearing in mind
that the spontaneous symmetry-breaking scale is $M_X$, which is fixed
for the particular model by the low-energy predictions, and then $\mu
= M_{X} /y_{m}$.

In this scheme, the semiclassical equations
for $\tilde{\phi}_0$ [15]
take the form (from now on,
the dot denotes differentiation with respect
to $\tau$)
\begin{eqnarray}
\ddot{y} & = & y \Bigr[\dot{\theta}^2 + \sin^2{\theta}
{}~\dot{\varphi}^{2}\Bigr] - 3
K \dot{y} \sqrt{\tilde{\rho}_{\phi}}
-{\partial \over \partial y}
{\widetilde V}^{(1)}(\tilde{\phi}) \; ,\\
\ddot{\theta} & = & - 2 {\dot{y} \over y} \dot{\theta} +\sin{\theta}
\cos{\theta} \; \dot{\varphi}^{2}  - 3
K \dot{\theta} \sqrt{\tilde{\rho}_{\phi}}
-{1\over y^2}{\partial \over \partial
\theta}{\widetilde V}^{(1)}(\tilde{\phi}) \; ,\\
\ddot{\varphi} & = & - 2 {\dot{y} \over y} \dot{\varphi} -
2 {\cos\theta \over \sin\theta} \dot{\varphi} - 3
K \dot{\varphi} \sqrt{\tilde{\rho}_{\phi}}
-{1 \over y^2 \sin^2{\theta} } {\partial \over \partial \varphi}
{\widetilde V}^{(1)}(\tilde{\phi}) \; .
\end{eqnarray}
With our notation, $K\equiv \sqrt{8 \pi G \mu^2/3{\cal G}^2}$,
and
\begin{equation}
{\tilde \rho}_{\phi} \equiv {1\over 2}
\biggr[{\dot y}^{2}+y^{2}\Bigr({\dot \theta}^{2}
+{\sin}^{2}\theta \; {\dot \varphi}^{2}\Bigr)\biggr]
+V^{(1)}(\phi) \; .
\end{equation}
Inflation eventually ends when, by virtue of decay mechanisms,
the energy stored in the scalar configuration is released to
lighter degrees of freedom (radiation). This is the so-called reheating
process, and the knowledge of the value of $\tau$, say $\tau_{f}$,
for which this actually happens, enables one to determine the
total $e$-fold number
$N(\tau_{f})$ of the inflationary model by solving the differential equation
\begin{equation}
\dot{N}(\tau) \equiv {d \over d\tau}
\ln\left({ a(\tau) \over a(0) }\right)
=K \sqrt{\tilde{\rho}_{\phi}}\; .
\end{equation}
A brief description of the reheating mechanism for our particular model
is presented in Sec. VII.

\section{Numerical analysis}

A necessary requirement to perform a numerical analysis of the
differential equations obtained in Sec. V, is a good knowledge
of the absolute minimum of the potential (5.2). As already stated in
Sec. III, if the inequality (3.12) is satisfied,
the effective potential is bounded from
below, but unfortunately, this does
not ensure that quantum corrections
to the tree-level potential will
make it possible to fix the value of the field
at the minimum, $y_{m}$, so that it is
sensibly different from zero. Thus, one has
to choose among the values of
$\alpha$, $\gamma$, $\delta$ and $\lambda$ satisfying (3.12), the ones
for which a symmetry breaking actually occurs [8].

In Tab. 1, we report for two
choices of the free parameters of the
classical potential (3.5) the corresponding values of $\theta_{m}$,
$\varphi_{m}$, $y_{m}$ and hence the values of $\mu$ and $\tilde{V}_0$.
The parameters $\alpha,\gamma,\delta$ and $\lambda$ have been chosen
in such a way that small couplings are achieved, and the minimum
of $y$ is of order 1.

In Fig. 1, we plot the solution for $y$
(obtained by means of the NAG Library routine D02CAF)
corresponding to the initial
conditions relevant for the $SU(4)_{\rm PS}\otimes SU(2)_{\rm L}
\otimes SU(2)_{\rm R}$ symmetry-breaking direction (see (3.4c)),
i.e. $\theta(0)=0,y(0)=10^{-5},{\dot \theta}(0)=0,
{\dot y}(0)=4. \times 10^{-8}$. Note that a vanishing value
of $\theta(0)$ and ${\dot \theta}(0)$ is the one resulting
from our particular choice of residual-symmetry direction, along
which the inflationary dynamics evolves. Moreover, a very small
value of $y(0)$, which approaches zero, reflects our choice
to start from a singlet state which has a complete $SO(10)$
symmetry (hence $y(0)=0$). Of course, a $y(0)$ value which
differs from zero is only taken for numerical convenience,
but our results are essentially independent of $y(0)$,
providing this is very small.
For numerical purposes, it has been thus
convenient to re-express the Eqs. (5.3)--(5.5) in terms of the
variables defined in Eqs. (A2)--(A4) of the Appendix. However, we keep
using the $y,\theta,\varphi$ parametrization in our paper, since it
makes it easier to describe symmetry breaking. What happens is that
$y$ characterizes the dynamics which is independent of the
symmetry-breaking direction,
whereas $\theta$ and $\varphi$ define the various
symmetry-breaking directions. After a slow-roll phase which is not
shown in Fig. 1 for the sake of clarity, $y$ starts increasing
until it reaches a region where it oscillates in the neighboorhood
of a relative minimum. The corresponding value of $\theta$ remains
equal to $0$. Interestingly, the time necessary to reach the region
of rapid oscillation for $y$ is of order $4. \times 10^{-35}$ sec,
and the corresponding $e$-fold number (see (5.7)) is $\simeq 100$.
However, such a large $e$-fold number is only achieved with the help
of a severe fine tuning of the ${\dot y}(0)$ value. This is indeed a
peculiar property of Coleman-Weinberg potentials, which have a
relative maximum at $y=0$. Further details about the oscillating
phase and the end of inflation will be given in Sec. VII.

In Fig. 2, $y$ is plotted against $\tau$ when the choice of
$\alpha,\gamma,\delta$ and $\lambda$ described on the second
line of numerical values of Tab. 1 is made.
In such a case, the following initial
conditions are chosen: $\theta(0)=0,y(0)=10^{-5},
\varphi(0)=\pi / 4, {\dot \theta}(0)=\sqrt{2} \times 10^{-3},
{\dot y}(0)=3. \times 10^{-8}, {\dot \varphi}(0)=0$.
In this case, however, $y$ oscillates in a neighboorhood of the
{\it absolute} minimum (cf. Fig. 1). The same happens for
$\theta$ and $\varphi$, whose evolution is not shown
for brevity. Moreover, a fine tuning of the initial
conditions leads again to an $e$-fold number of order 100, and the
duration of the inflationary stage is the same as in Fig. 1.

The cases described in Figs. 1 and 2 are indeed also relevant
for the analysis of the reheating stage of the early universe,
as shown in Sec. VII, where a brief outline of such a phenomenon
is presented.

\section{Reheating}

When the energy stored in the scalar-field configuration
is released to the relativistic degrees of freedom (hereafter $\rho_R$
denotes the corresponding energy density) and begins to dominate
the total energy of the universe, the evolution undergoes the
so-called reheating phase. Such a phase, which is the necessary last stage
of the inflationary dynamics,
is as important as the exponential expansion itself, since it
ensures the end of the exponential growth and the reheating of the
universe. In fact, if the radiation energy dominates the total energy
of the system, through the equation of state, the energy density and
the pressure $p_{R}$ satisfy the condition $\rho_R+3 p_R = 4
\rho_R>0$, which implies $\ddot{a}<0$.

According to the recent models of reheating, this process
can be divided into three stages [16--18]. In the first stage, the
scalar field $\phi$ decays into massive bosons via parametric
resonance. The second stage consists in the decay of previously
produced particles, and the last stage leads to thermalization
(for our purposes, only the first two stages will be considered).

As far as the parametric resonance is concerned, it can be studied
starting from the equation for quantum fluctuations of a scalar field
$\omega$, quadratically coupled to $\phi$. As shown in Ref. [16], on
setting $z=m_{\phi}t$, such an equation may be cast in the form
\begin{equation}
\omega_{k}''+\Bigr[A(k)-2q \cos(2z)\Bigr]\omega_{k}=0 \; ,
\end{equation}
where the {\it prime} denotes differentiation with respect to $z$.
Moreover, $A(k)=k^{2}/m_{\phi}^{2}a^{2}+2q$ (${\vec k}/a$ being
the physical momentum), and $q=g^{2}\Phi^{2}/4m_{\phi}^{2}$. In
the model considered in Refs. [16--18], $\Phi$ is the amplitude of
oscillations of the field $\phi$, and $g$ is a small coupling
constant. Interestingly, an exponential instability of the
solutions exists and it can be interpreted
as a rich particle production.
This phenomenon is best tackled by studying the
stability/instability chart of the Mathieu equation [16--18].

In particular, the field $\omega$ may be given by the fluctuations
of the scalar field $\phi$ itself. In such a case, one has to
start from Eq. (A9) of the appendix, where the effect of
$3H{d {\delta z}\over dt}$ is neglected, following Ref. [16].
Parametric resonance is only achieved if ${\widetilde M}(z_{1}^{0},
z_{2}^{0}, z_{3}^{0})$ therein is positive for some values
of $\alpha,\gamma,\delta,\lambda$,
in the neighboorhood of which the scalar field starts
oscillating. In Figs. 3 and 4, we plot
the function $\widetilde M$ against $\theta$ and $\varphi$,
for the values of $\alpha,\gamma,\delta$ and $\lambda$ reported
on the first and second line of
numerical values of Tab. 1, respectively. In each
figure, the closed curve represents those values of $\theta$
and $\varphi$ for which $\widetilde M$ vanishes. Within the
region bounded by such a curve, $\widetilde M$ takes negative
values, whereas it is positive outside.

Interestingly, in Fig. 3, $\theta_{m}$ and $\varphi_{m}$ lie
in the region of parametric resonance, i.e., outside the region
where $\widetilde M <0$, while the converse holds in Fig. 4.
Of course, since Fig. 1 corresponds to a case when $\theta$
remains equal to zero, parametric resonance is indeed achieved
because, for this particular choice of parameters,
${\widetilde M}(y,\theta=0,\varphi)$ is positive, as
shown in Fig. 3.

Thus, our analysis shows that the inflationary dynamics with
residual symmetry $SU(4)_{PS} \otimes SU(2)_{L} \otimes
SU(2)_{R}$ (see (3.4c)) leads to slow roll and also to parametric
resonance. These properties add evidence in favour of SO(10)
GUT models being physically relevant, although one should
bear in mind that fine tuning problems remain. However, the case
described by Fig. 4 is not, by itself, ruled out. One has
instead to consider a scalar field $\omega$ which is {\it not}
given by the fluctuations of $\phi$ itself [16--18].

We now briefly consider the second stage of the reheating
process. Once that the quantum
fluctuations of the $\underline {210}$ IRR have been produced
via parametric resonance, they have to decay into relativistic
degrees of freedom, whose mass is negligible with respect to
the $M_{X}$ scale. Note that the decay of the
$\underline {210}$ representation into a fermion-antifermion
pair cannot occur, since it makes it necessary to introduce a
Yukawa coupling of the $\underline {210}$ to fermions, which
would lead in turn to an undesirable mass
term for fermions of order $M_{X}$.
On similar ground, any gauge boson coupled to
$\underline {210}$ would acquire mass at the scale $M_{X}$,
and hence such a coupling cannot lead to decays into
relativistic degrees of freedom. One is thus left with decays
into other Higgs particles of smaller mass. Indeed,
the presence of lighter Higgs fields in GUT theories is particularly
evident in $SO(10)$, where the spontaneous symmetry
breaking which leads to the standard electroweak model occurs in two
steps at different mass scales: i.e.,
\begin{equation}
SO(10)~ {\mathrel{\mathop{\longrightarrow}^{M_{X}}}}~ G ~
{\mathrel{\mathop{\longrightarrow}^{M_{R}}}}~
SU(3)_C \otimes SU(2)_L \otimes U(1)_Y \; ,
\end{equation}
where $M_X$ is of order $10^{15}-10^{16}~GeV$ to be compatible with
the lower limit on proton decay,
$G$ is one of the intermediate symmetry
groups appearing in Eqs. (3.4a)--(3.4f),
and the scale $M_{R}$ is the one
relevant for neutrino physics. Note that every breaking phase in (7.2)
is mediated by a different scalar field belonging to a IRR of
$SO(10)$. A realistic model can be constructed for example by
considering in addition to the $\underline{210}$ ($\phi$), the scalar
fields ($\psi$, $\bar{\psi}$) belonging to the representation
$\underline{126} \oplus \overline{\underline{126}}$, and two
ten-dimensional irreducible representations $(\chi)$ which mediate the
electroweak symmetry breaking [19]
\begin{equation}
SO(10) ~{\mathrel{\mathop{\longrightarrow}^{<\phi>}}}~ G~
{\mathrel{\mathop{\longrightarrow}^{<\psi>}}}~
SU(3)_C \otimes SU(2)_L \otimes U(1)_Y~
{\mathrel{\mathop{\longrightarrow}^{<\chi>}}}~
SU(3)_C \otimes U(1)_Q\; .
\end{equation}
In this model, to obtain the above symmetry-breaking pattern, one has
to replace the simple tree-level potential of Eq. (3.2) by a much more
complicated expression $V(\phi,\psi,\bar{\psi},\chi)$ (see Eqs.
(7)--(11) of Ref. [19]).
Still, as far as the inflation at the highest
energy scale is concerned, only the terms of the potential containing
uniquely the ${\underline {210}}$ representation with the
associated scalars are relevant. This is
why we neglected in (3.2) the contributions resulting from the other
representations. The other terms, however, are important when the
second stage of reheating is considered,
since they can mediate the decay of the
massive Higgs in the lighter $\psi$, $\bar{\psi}$ and $\chi$. The
interacting terms between different scalar representations
are the ones actually relevant in
that they do not provide mass for light particles,
while decay processes are allowed.
Such interacting terms occur in quartic form,
involve the whole set of representations of $SO(10)$, and hence,
being linear in $\phi$, do not provide mass for the lighter
Higgs particles (Eq. (11) of Ref. [19]).

\section{Concluding remarks}

Our paper has studied the relevance for inflationary cosmology
of a nonsupersymmetric GUT theory which is consistent with
the available data on nucleon lifetimes. This is the $SO(10)$
model in the 210-dimensional irreducible representation
(see Refs. [9,20] and literature cited therein).

Starting from a quartic tree-level potential in the case of
minimal coupling, we have studied the flat-space limit of
the semiclassical field equations, where the one-loop effective
potential contains the logarithmic terms which result from the
Coleman-Weinberg method for the evaluation of radiative
corrections [5]. Following Ref. [8], when the Higgs scalar field
belongs to the 210-dimensional irreducible representation
of $SO(10)$ (this corresponds to the highest energy levels),
attention has been restricted to the mass matrix relevant
for the $SU(3)_{C} \otimes SU(2)_{L}
\otimes U(1)_{Y}$ symmetry-breaking
direction, to agree with the low-energy phenomenology of
the standard model of particle physics.

The main result of our investigation is that the inflationary
dynamics with residual symmetry $SU(4)_{PS}\otimes
SU(2)_{L} \otimes SU(2)_{R}$ leads to a sufficiently long
inflationary stage, and then a reheating process occurs
via parametric resonance and subsequent decay of the particles
produced previously. Such processes are three-body decays of
$\underline {210}$ into the
massless components of the $\underline {126}$,
$\overline{\underline {126}}$ and $\underline {10}$
representations, and result from the mutual quartic coupling
of all irreducible representations. The intermediate symmetry
$SU(4)_{PS} \otimes SU(2)_{L} \otimes SU(2)_{R}$ has been
extensively studied in the literature on particle physics,
since it occurs in the most promising $SO(10)$ models [9,20].
Thus, a natural and deep link
between low-energy phenomenology, grand unification,
inflationary cosmology and physical processes in the very early
universe seems to emerge from our work.

Of course, we have only studied some {\it particular} values
of the parameters of the tree-level potential. Although
particle physics and cosmology lead to restrictions on such
parameters, our choices are by no means exhaustive. It now
appears interesting to get a more quantitative understanding of the
reheating process outlined in Sec. VII. Moreover, from the
point of view of perturbative properties of quantum field
theory, it is necessary to study in detail the nonlocal
contributions to the semiclassical field equations. These
arise already in scalar electrodynamics [13], by virtue of
the coupling of the scalar field to the gravitational
background, and are receiving careful consideration since
they are related to dissipative and nondissipative
phenomena in the early universe [13,21].
It also appears interesting to study the corrections to
the effective potential resulting from a suitable
generalization to $SO(10)$ models of the technique
described in Ref. [22].

\acknowledgements
We are much indebted to Franco Buccella
for teaching us all what we know
about $SO(10)$ models, and to Gianpiero Mangano, Ofelia Pisanti
and Luigi Rosa for several useful conversations
about the research described in this publication.
\appendix
\section*{}
To obtain the equation for the quantum fluctuations of the
$\underline{210}$ IRR, we follow the notation of Ref. [8], for
which the most general singlet $\phi_0$
with respect to the gauge group
$SU(3)_{C} \otimes SU(2)_{L} \otimes U(1)_{Y}$
is written down in Eq. (3.3).
Within this framework, the Lagrangian
for scalar fields only, which are assumed to be
spatially homogeneuos, reads
\begin{equation}
{\cal L}_{\phi_0} = a^{3} \left[
{ 1\over 2} \left({d z_1 \over dt}\right)^2 +
{ 1\over 2} \left({d z_2 \over dt}\right)^2 +
{ 1\over 2} \left({d z_3 \over dt}\right)^2
\right]
+\sum_{l=1}^{3}g^{ij}z_{l,i}z_{l,j}
- V(z_1,z_2,z_3) \; ,
\end{equation}
where there is summation over the repeated indices $i$ and $j$.
We now take the one-loop expansion of (A1) by writing
$z_{i}=z_{i}^{0}+{\delta z}_{i}$, and we make the ansatz
$\delta z_{i}=\delta z$ for simplicity.
The three independent
degrees of freedom of the scalar field turn out to be
\begin{eqnarray}
z_{1}^{0} & = & \| \phi_{0} \| \sin\theta \cos\varphi \; , \\
z_{2}^{0} & = & \| \phi_{0} \| \sin\theta \sin\varphi \; ,  \\
z_{3}^{0} & = & \| \phi_{0} \| \cos\theta \; .
\end{eqnarray}
Thus, the tree-level potential (3.5) can be
re-written in terms of (A2)--(A4) by pointing out that
\begin{eqnarray}
\|\phi_0 \|^2 & = & (z_{1}^{0})^2
+(z_{2}^{0})^{2}+(z_{3}^{0})^2\; ,\\
f_{\alpha} \|\phi_0 \|^4 &\equiv&
{\Bigr((z_{1}^{0})^{2}+(z_{2}^{0})^{2}\Bigr)}^{2}
+(z_{2}^{0})^{2}{\Bigr(2z_{1}^{0}
+\sqrt{3}z_{3}^{0}\Bigr)}^{2}
+{3\over 4}(z_{2}^{0})^{4}\; ,\\
f_{\gamma} \|\phi_{0} \|^{4}
&\equiv& {\biggr(z_{1}^{0}z_{3}^{0}
+{(z_{2}^{0})^{2}\over \sqrt{3}}
\biggr)}^{2}
+(z_{1}^{0}z_{2}^{0})^{2}+f_{\alpha} \|\phi_0 \|^4\; ,\\
f_{\delta}\|\phi_{0} \|^{4} & \equiv&
\Bigr(30f_{\gamma}-25f_{\alpha}\Bigr) \|\phi_{0} \|^{4}
+{\biggr(2(z_{1}^{0})^{2}-{(z_{2}^{0})^{2}\over 2}
-3(z_{3}^{0})^{2}\biggr)}^{2}\;.
\end{eqnarray}
One thus gets the
following equation for the quantum fluctuation:
\begin{equation}
{d^2 \delta z \over d t^2} + 3 H {d \delta z \over d t}
+ \biggr[{k^{2}\over a^{2}}
+{\widetilde M}(z_{1}^{0},z_{2}^{0},z_{3}^{0})\biggr]
\delta z = 0 \; ,
\end{equation}
where $k \equiv \sqrt{{\vec k}^{2}}$,
${\widetilde M}(z_{1}^{0},z_{2}^{0},z_{3}^{0})$
is defined as
\begin{equation}
{\widetilde M}(z_{1}^{0},z_{2}^{0},z_{3}^{0}) \equiv
{2 \over 3}
\left(c_{110}~ z_{1}^{0} z_{2}^{0}
+ c_{101}~ z_{1}^{0} z_{3}^{0} + c_{011}~ z_{2}^{0} z_{3}^{0}
+c_{200}~(z_{1}^{0})^{2}
+ c_{020}~(z_{2}^{0})^{2} + c_{002}~(z_{3}^{0})^{2} \right) \; ,
\end{equation}
and the $c_{abd}$ coefficients are given by
\begin{eqnarray}
c_{110} & \equiv & (3+ \sqrt{3}) ~\alpha
+7\left( 1 + {1 \over \sqrt{3}} \right) \gamma
+ {80 \over 9}(2+ \sqrt{3})~\delta
+8\lambda \; , \\
c_{101} & \equiv & {\sqrt{3} \over 2} \alpha
+\left(1 + {7 \over 2 \sqrt{3}} \right) \gamma
+ {40 \over 9}\sqrt{3}~\delta
+ 8\lambda \; , \\
c_{011} & \equiv & \left( { 3 \over 2} + \sqrt{3} \right) \alpha
+ \left(3 + {7 \over \sqrt{3}} \right) \gamma
+{80 \over 9}\sqrt{3}~\delta
+8\lambda \; , \\
c_{200} & \equiv & {3 \over 2} \alpha
+ {7 \over 2} \gamma
+ {40 \over 9} \delta
+10 \lambda \; , \\
c_{020} & \equiv & { 1 \over 2}
\left( {39 \over 8} +\sqrt{3} \right)\alpha
+ {1 \over 2}\left({45 \over 4} +
{7 \over \sqrt{3}} \right) \gamma
+{20 \over 9} (5 + 2 \sqrt{3}) ~\delta
+ 10 \lambda \; , \\
c_{002} & \equiv & { 3 \over 8} \alpha + \gamma + 10 \lambda \; .
\end{eqnarray}

\newpage
\bigskip\bigskip
\par\noindent
{\bf Table 1.}

\vspace{.5cm}

\noindent
\begin{tabular}{|c|c|c|c|c|c|c|c|c|}
\hline
& & & & & & & &\\
$\alpha$ & $\gamma$ & $\delta$ & $\lambda$ & $\theta_m$ & $\varphi_m$ &
$y_m$ & $\mu$ & ${\widetilde V}_{0}$\\
& & & & & & & $(10^{16}~GeV)$ & \\
\hline
& & & & & & & & \\
$-3. \times 10^{-2}$
& $5. \times 10^{-5}$
& $-5. \times 10^{-6}$
& $1. \times 10^{-2}$
& $1.31$
& $0.94$
& $1.68$
& $0.34$
& $2.49 \times 10^{-2}$\\
$ $ & $ $ & $ $ & $ $ & $ $ & $ $ & $ $ & $ $ & \\
\hline
& & & & & & & & \\
$-1. \times 10^{-4}$
& $5. \times 10^{-7}$
& $-5. \times 10^{-8}$
& $3.39 \times 10^{-5}$
& $1.70$
& $1.35$
& $2.09$
& $0.27$
& $3.50 \times 10^{-2}$\\
$ $ & $ $ & $ $ & $ $ & $ $ & $ $ & $ $ & $ $ & \\
\hline
\end{tabular}
\vskip 2cm
\noindent
{Figure captions:}
\vskip 0.3cm
\noindent
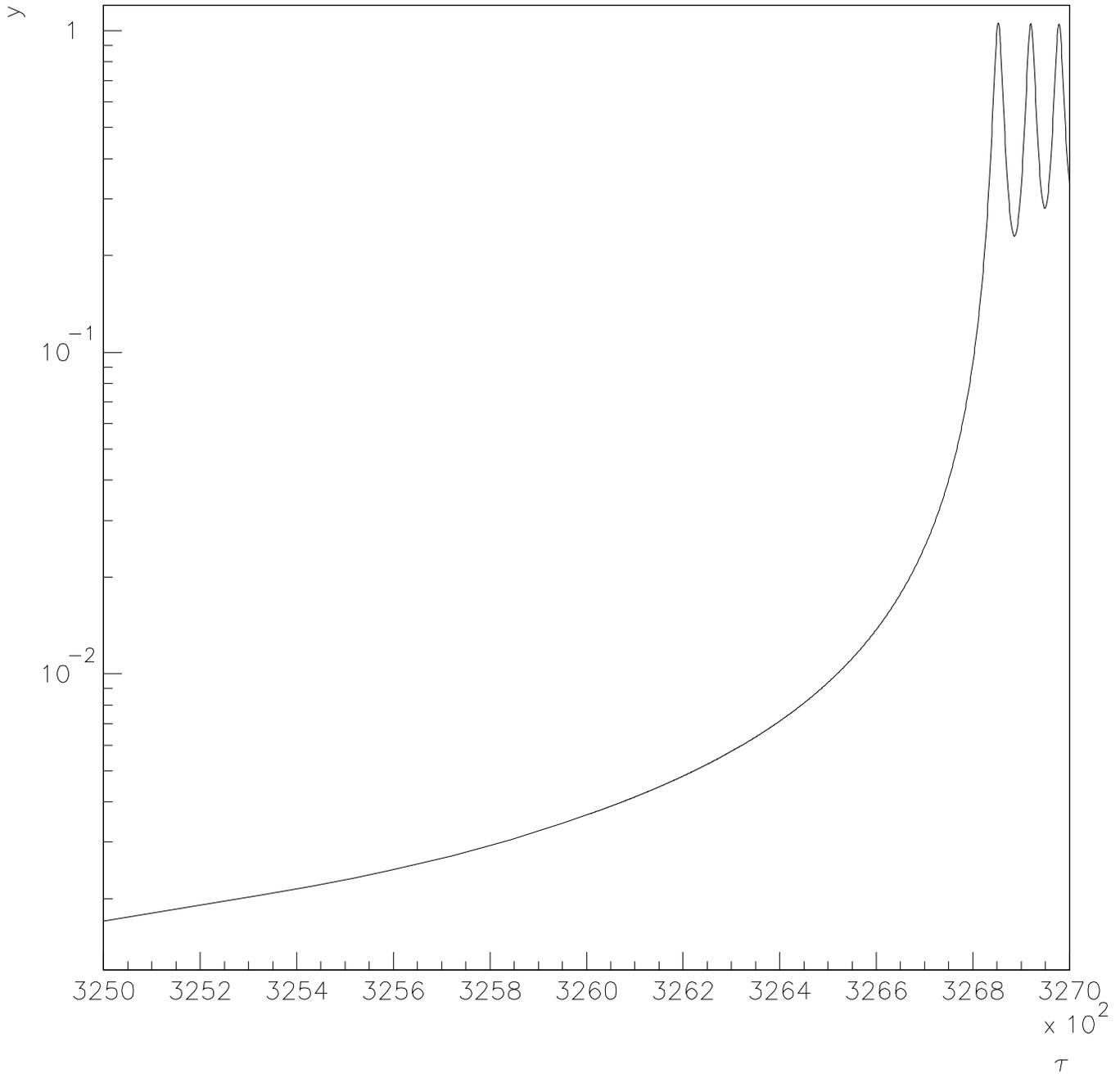
FIG. 1. The solution of Eq. (5.3) is plotted against $\tau$,
for the first set of values of $\alpha,\gamma,\delta,\lambda$ shown
in Table 1.
\vskip 0.3cm
\noindent
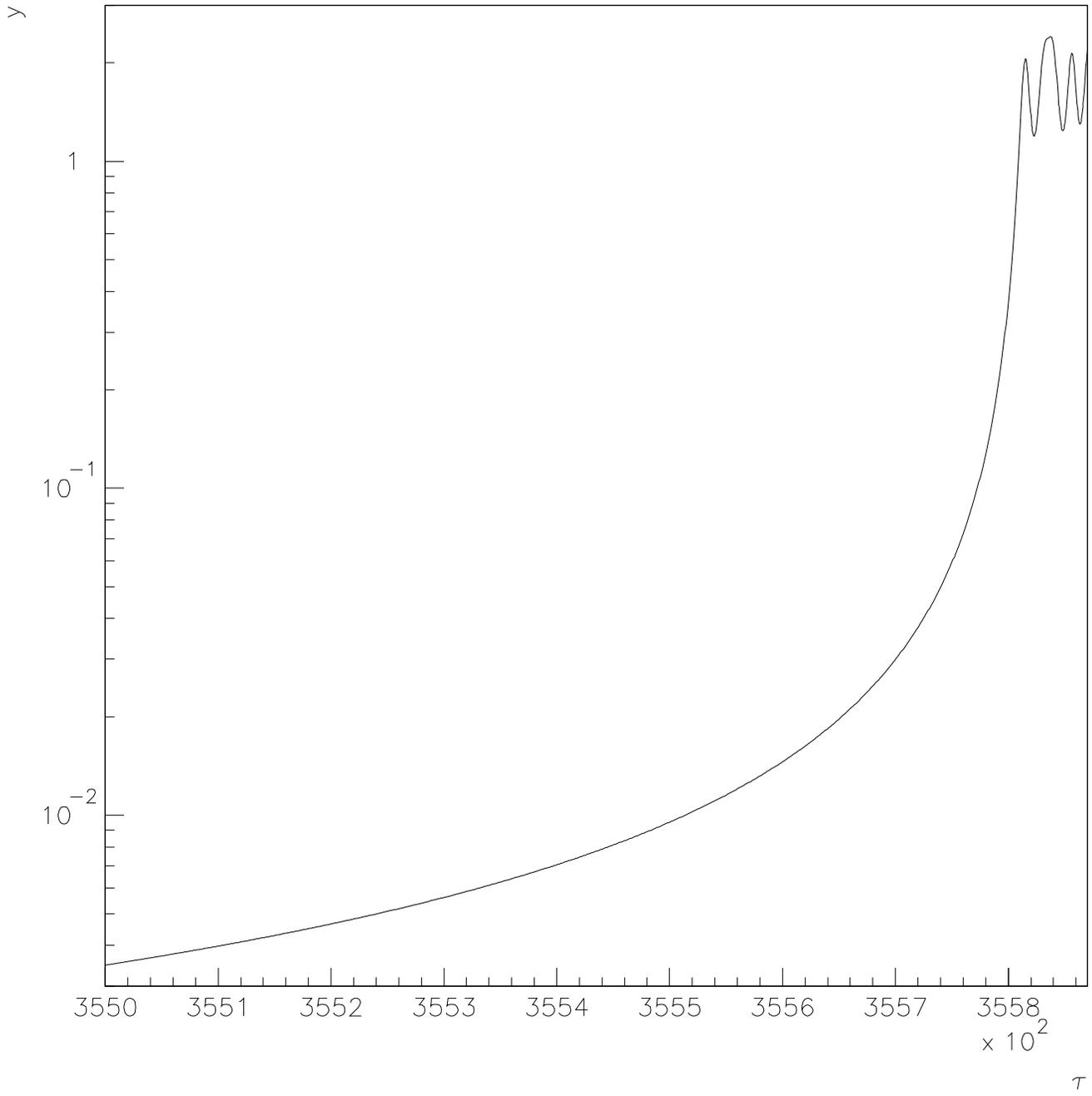
FIG. 2. The solution of Eq. (5.3) is plotted against $\tau$,
for the second set of values of $\alpha,\gamma,\delta,\lambda$
shown in Table 1.
\vskip 0.3cm
\noindent
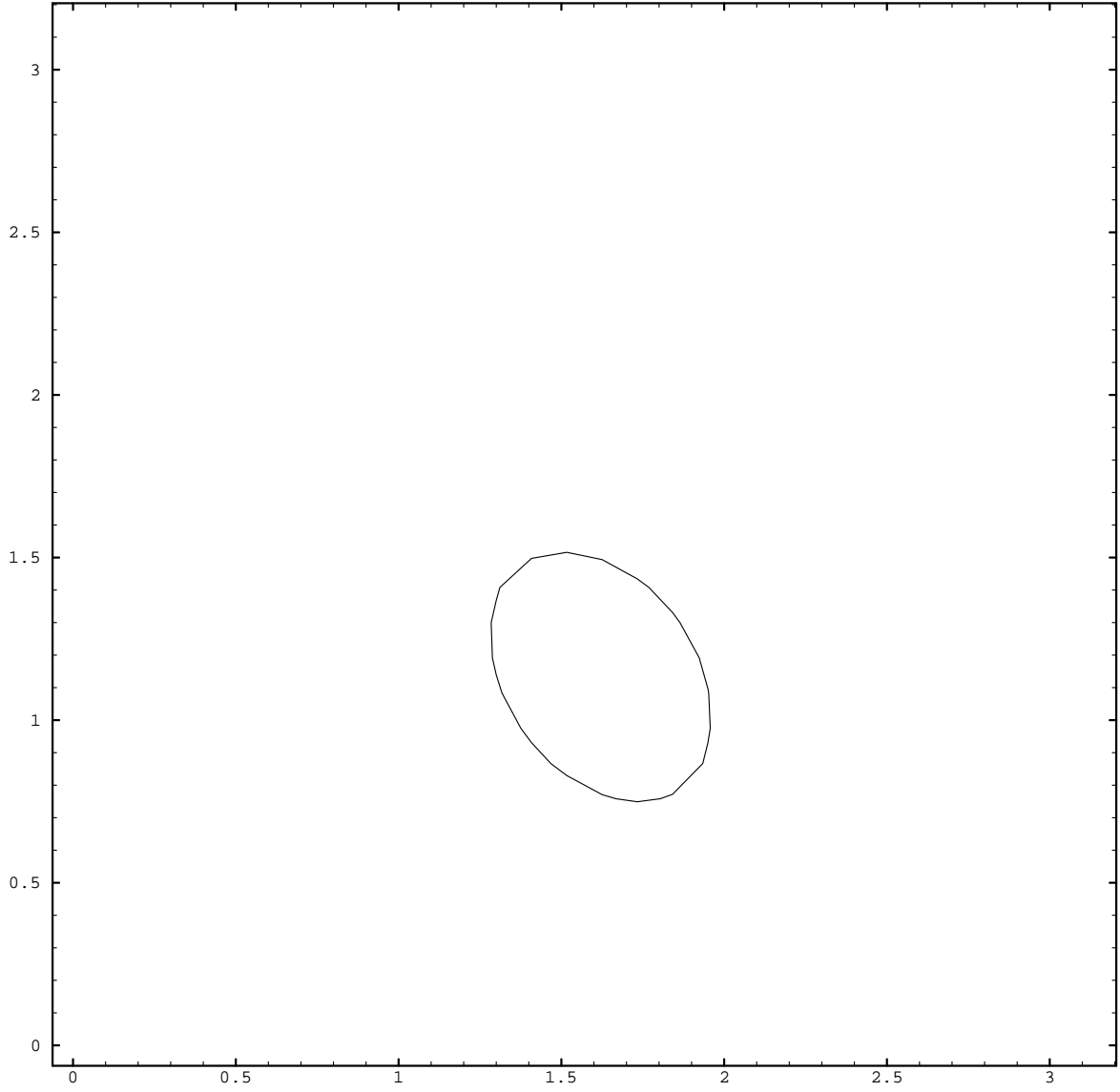
FIG. 3. The function defined in Eq. (A10),
divided by $y^{2}$, is plotted against
$\theta$ and $\varphi$, for the first set of values of
$\alpha,\gamma,\delta,\lambda$ given in Table 1.
\vskip 0.3cm
\noindent
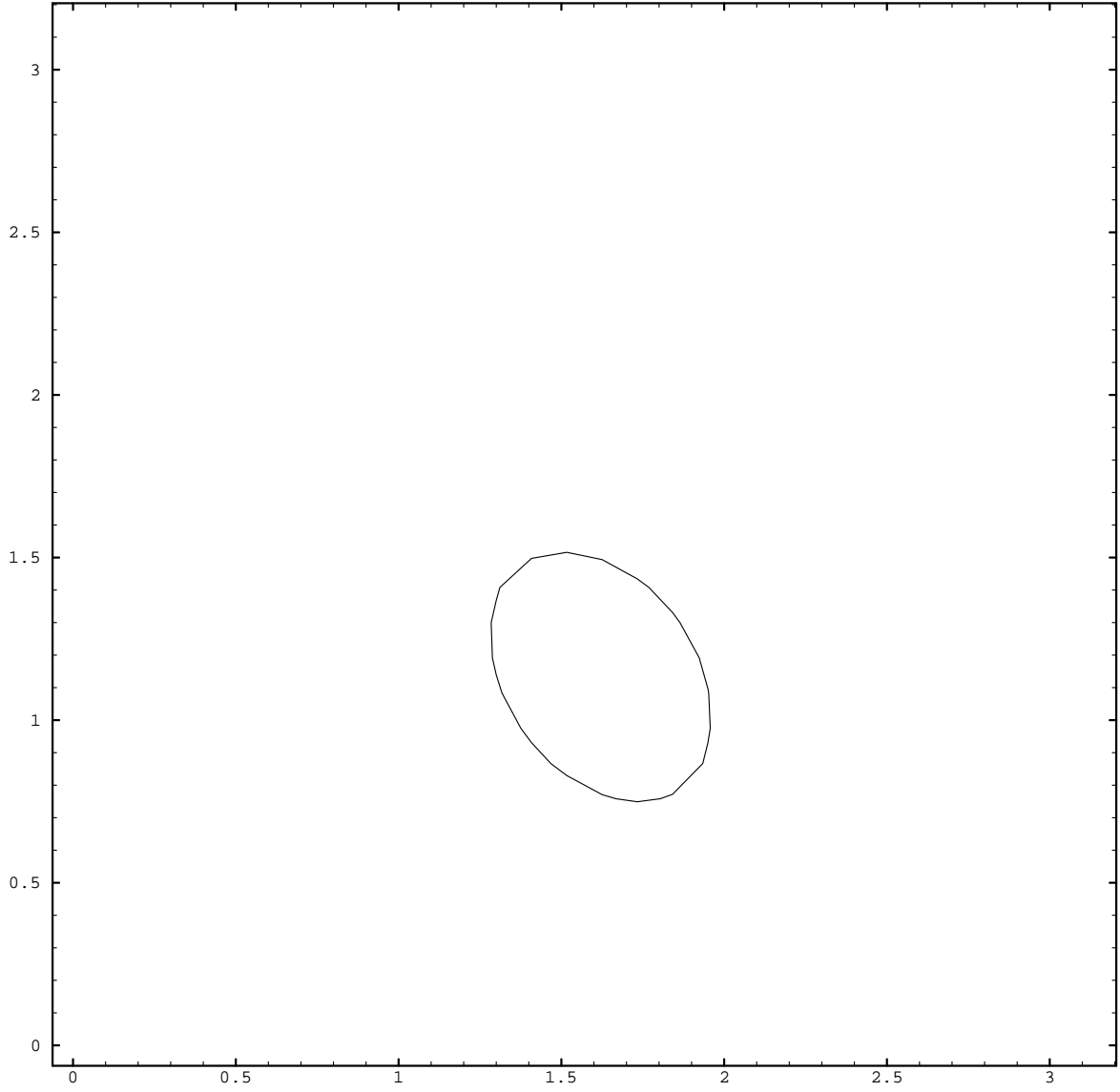
FIG. 4. The function defined in Eq. (A10),
divided by $y^{2}$, is plotted against
$\theta$ and $\varphi$, for the second set of values of
$\alpha,\gamma,\delta,\lambda$ given in Table 1.
\end{document}